\documentclass[twoside,11pt]{LCWS13}
\usepackage[latin1]{inputenc}
\usepackage[pdftex]{graphicx,epsfig,color}
\usepackage{wrapfig,rotating}
\usepackage{amssymb,amsmath,array}
\usepackage{cite}
\usepackage{multirow}
\usepackage{lineno}
\usepackage{caption}
\usepackage{subcaption}
\usepackage{watermark}
\usepackage{floatflt}
\usepackage{epstopdf}

\DeclareGraphicsExtensions{.pdf,.png,.jpg,.eps}

\def\logo{\hfill\includegraphics[width=0.2\textwidth]{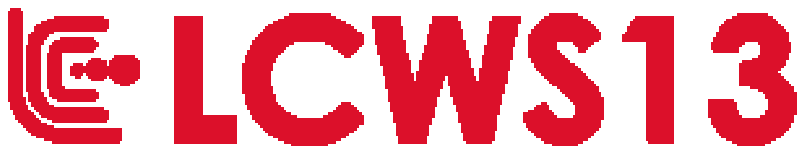}}

\newcommand{\ecalRinn}{\ensuremath{R^{\mathrm{inner}}_{\mathrm{ECAL}}}}
\newcommand{\tpcOutRad}{\ensuremath{R^{\mathrm{outer}}_{\mathrm{TPC}}}}

\begin{document}
\title{
    \logo\\
    \vspace{-5mm}\noindent{\textcolor{black}{\rule{1.05\textwidth}{0.4pt}}}\vspace{10mm}\\
ILD SiW ECAL and sDHCAL dimension-performance optimisation\\
{\vspace{3mm}\it\normalsize Talk presented at the International Workshop on Future Linear Colliders (LCWS13)\\
    \vspace{-2mm} Tokyo, Japan, 11-15 November 2013.\vspace{5mm}}
} 
\author{Trong Hieu TRAN
\vspace{.3cm}\\
Laboratoire Leprince-Ringuet, Ecole polytechnique, CNRS/IN2P3\\
F-91128 Palaiseau, France\\
}

\maketitle

\begin{abstract}
    The ILD, International Large Detector, is one of the detector concepts for a future linear
    collider. Its performance is investigated using Monte-Carlo full simulation and PandoraPFA.  
    Among several options, a combination of the silicon-tungsten electromagnetic calorimeter (SiW ECAL)
    and the semi-digital hadronic calorimeter (sDHCAL) presenting the highest granularity calorimeters,
    is here investigated.
    It is shown that by reducing the
    radius and length of the entire detector by a factor of $\sim1.3$ with respect to the
    baseline dimensions, the jet energy resolution 
    is degraded by~8 to 19\% in the range of~45 and~250~GeV. 
    The price of ILD which scales roughly quadratically with the ILD dimensions may be reduced by
    a factor of nearly two.
    A similar study made with the SiW ECAL and the analog hadronic calorimeter (AHCAL)
    shows that for an inner radius of ECAL of about~1.4~m, the
    performance is comparable between sDHCAL and AHCAL.
\end{abstract}

\section{Introduction}
The International Large Detector (ILD) concept~\cite{bib:ildLOI, bib:ilcTDR}, combining an excellent tracking and a fine granular calorimetry system, has been optimised for particle flow approach (PFA)~\cite{bib:pfa1,bib:pfa2}.
The calorimetry system is composed of an electromagnetic calorimeter (ECAL) and a hadronic calorimeter (HCAL). The ECAL consists of interleaved layers of absorbing material (tungsten) whereas the HCAL uses steel as absorber. There are two candidates for active material for both ECAL and HCAL. For ECAL one considers to use silicon sensors or scintillator plastic strips. For the HCAL, scintillator tiles and gaseous devices are proposed, respectively readout in analogue and semi-digital modes,
the so-called AHCAL and sDHCAL.

A first estimation of cost of the ILD has been presented in the letter of intent for ILD~\cite{bib:ildLOI} . 
It has been carefully revisited in the detailed baseline design~\cite{bib:ilcTDR}.
In the latter it was estimated that the total cost of ILD can reach $\sim400$~MILCU.
\footnote{ILCU is a common unit for ILC cost estimation, which is defined as 1~United State official currency, USD, on 1 January 2007.}
The cost estimation for individual components of the full detector has shown that
the most expensive parts are the return yoke and the SiW ECAL. 

During the last few years, many studies have been performed on the optimisation of the cost-performance.
For the SiW ECAL the cost is proportional to the silicon surface which is of about 2400~m$^2$ in the baseline design of ILD.
In a recent study we have shown a possibility of reducing the number of layers 
(for a fixed number of radiation length, 23$X_0$), cf.~\cite{bib:ilcTDR}.

We consider here the feasibility of reducing the whole ILD size, i.e. radius and length proportionally.
A similar study have been done for the ILD as a combination of SiW ECAL and AHCAL, cf.~\cite{bib:pfa2}.
We concentrate only on the option of using silicon sensors for ECAL (SiW ECAL) and gas as active materials for HCAL (sDHCAL).
For the SiW ECAL, a granularity of $5\times5$~mm$^2$ is chosen for the silicon sensor size.
The gaseous detector HCAL is based on Glass Resistive Plate Chamber (GRPC) technology.
The option of $1\times1$~cm$^2$ cells was chosen in a study on full ILD
detector model is a compromise between a few mm cell size a better energy resolution 
provided by 
and the handling (and cost) of a huge number of electronic readout channels.
The sDHCAL is readout in semi-digital mode, where the energy information is coded in 2-bit 
corresponding to three thresholds. This helps to correct the saturation effect and improve the energy resolution at high energies.

The list of softwares which are used for the analysis chain is listed in section~\ref{sec:simrec} 
together with the simulation, calibration and jet energy reconstruction procedures. 
Section~\ref{sec:perf.jer} discusses the ILD performance in terms of jet energy resolution.
    Some cross-checks are shown in section~\ref{sec:xchecks}.
    And finally, all the results are summarized in section~\ref{sec:conclusions}.

\section{Simulation and reconstruction}
\label{sec:simrec}
\subsection{Simulation}
For the simulation, we use Mokka~\cite{bib:mokka}, an overlayer of GEANT4 version 9.5~\cite{bib:geant4},
allowing for a complete parametrised geometry of ILD.
The process
\begin{equation}
   e^+ + e^- \rightarrow Z\rightarrow q\bar{q}
\end{equation}
where $Z$ is generated at rest at 
different centre-of-mass energies, 91, 200, 360 and 500~GeV is used. Only light quarks 
($q$ = $u$, $d$, $s$) are considered for the time being.
For the modelling of hadronic showers, the Geant4 physics list \verb QGSP_BERT \ has been used. This physics list is based on the Precompound model of nuclear evaporation~\cite{bib:geant4doc} (QGSP) for
high energy interactions, and the Bertini (BERT) cascade model~\cite{bib:bert}
for intermediate energy interactions.

The radius and the length of the detector are changed proportionally such that their ratio is kept
like in the baseline design.
The modelled ECAL inner radii ($R^{\mathrm{inner}}_{\mathrm{ECAL}}$) and barrel lengths are given in table~\ref{tab:RadLen} 
together with the corresponding outer radius of the Time Projection Chamber (TPC), $\tpcOutRad$.
\begin{table}
   \centering
   \begin{tabular}{rcccc}
      \hline
      $R_{\mathrm{ECAL}}^{\mathrm{inner}}$ (mm) & 1843 & 1600 & 1400 & 1200 \\
      $R_{\mathrm{TPC}}^{\mathrm{outer}}$ (mm)  & 1808 & 1565 & 1365 & 1165 \\
      half length (mm)  & 2350 & 2040 & 1785 & 1530 \\\hline
   \end{tabular}
   \caption{Parameters for different ILD configuration.}
   \label{tab:RadLen}
\end{table}
All other parameters of ILD are kept unchanged, e.g. ECAL, sDHCAL segmentation, thicknesses ($23X_0$ and $6\lambda_I$), cell size.

Events are reconstructed within the standard ILC Marlin~\cite{bib:marlin} framework, version 01-16. 
The jets are reconstructed by the Pandora particle flow reconstruction algorithms, version 00-09, which was first developped in~\cite{bib:pfa2, bib:pfa3} and its recent updates and improvements are presented in~\cite{bib:pandora2}.

\subsection{Calibration}
To achieve best results with Pandora, it should be recalibrated for every ECAL inner radii.
The calibration procedure is carried out in the following steps:
\begin{itemize}
   \item determination of ECAL and HCAL hit conversion factors between deposited charge and energy; 
      the factors for ECAL are determined using $\gamma$'s at 10~GeV, whereas for the sDHCAL
      the parameters have been determined for sDHCAL prototype in semi-digital mode using
      pion's of energies in the range of~5 and~80~GeV, cf.~\cite{bib:sdhcalcalib} for more details;
   \item hadron calibration: the initial ECAL and HCAL calibration parameters in the first step are determined
      for single particles. However, as hadrons in jets are a mixture of $K_L$'s, neutrons and 
      anti-neutrons, 
      these parameters need to be retuned to minimise the Monte-Carlo jet energy
      resolution. Two scaling parameters are defined: the weights which should be applied to
      the energy deposits in the ECAL and in the HCAL being identified as part of the hadronic
      showers. The minimisation is done through a scanning procedure by changing these two weights
      in order to search for a minimum of jet energy resolution.
      It is found that the ECAL energy weight should be 1.05 and 1.25 for the HCAL weight
      for all of ILD setups in this study;
   \item MIP calibration: determine the conversion factor from calorimeter hit to a minimum
      ionising particle (MIP) equivalent. This calibration step is done using $\mu^-$'s at 10~GeV;
   \item angular correction: the energy lost in the gaps between modules in the barrel regions, and between module parts are currently not taken into account by reconstruction 
      software. It is neccessary to do a correction to compensate the lost energy. For the gaseous sDHCAL the angular correction also account for the varyiing sampling fraction.
      The correction is determined separately for $\gamma$'s and $K^{0}_L$'s by
      adjusting the mean value of the PandoraPFA reconstructed energy distribution for each bin of $\left|\cos\theta_{q}\right|$.
      Then it is applied to photons' and neutral particles in jets. 
      An improvement of the resolution by about~2\% in most of cases was observed, compared to
      the jet energy resolution without correction.
\end{itemize}


\section{Jet energy resolution}
\label{sec:perf.jer}
Jet energy resolution (JER) is a key estimator of the performance of ILD since it is related to the 
separation of hadronic decays of $W$ and $Z$ bosons through the reconstruction of their 
di-jet invariant masses.
The di-jet energy, $E_{jj}$, is then summed up from single particle energies.
The jet energy resolution is estimated by RMS90 method, which is the root mean square of
the smallest energy range containing 90\% of the events.
The single jet energy resolution (SJER) is then defined as:
\begin{equation}
    \frac{\mathrm{RMS_{90}}(E_j)}{\mathrm{mean}_{90}(E_j)} = \frac{\mathrm{RMS_{90}}(E_{jj})}{\mathrm{mean}_{90}(E_{jj})}\times\sqrt{2}
\end{equation}
where the RMS$_{90}$($E_{jj}$) and the mean$_{90}$($E_{jj}$) are the RMS and the mean value of the
distribution of dijet energy, $E_{jj}$. These are
calculated from the total reconstructed energy distribution. To avoid the gap between the barrel and the endcap regions of the detector, only jets in the barrel region are considered.\\
\begin{figure}
    \centering
    \begin{subfigure}[t]{0.48\textwidth}
        \includegraphics[width=\textwidth]{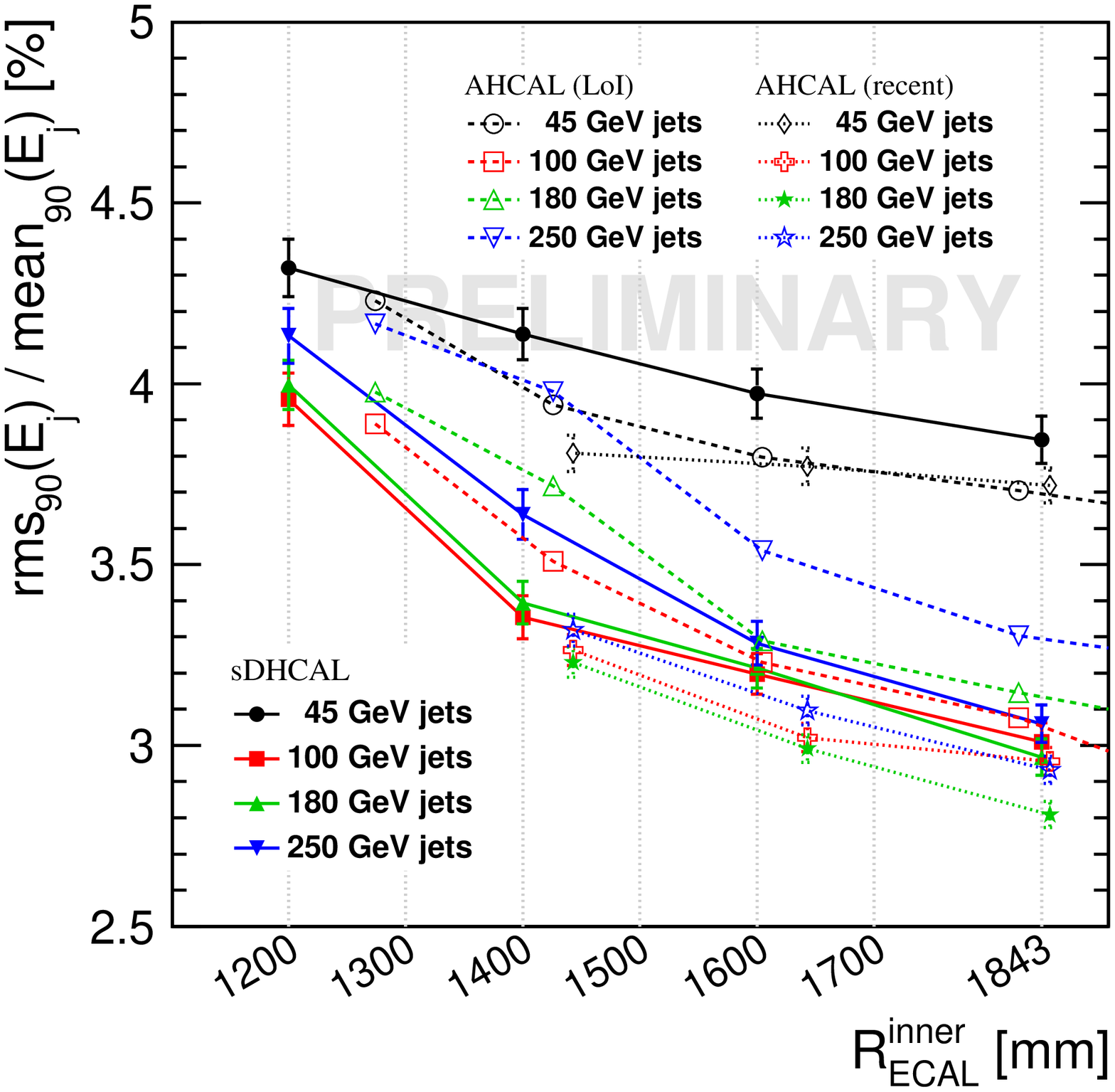}
        \caption{}
        \label{Fig:JERvsRad}
    \end{subfigure}
    \quad
    \begin{subfigure}[t]{0.48\textwidth}
        \includegraphics[width=\textwidth]{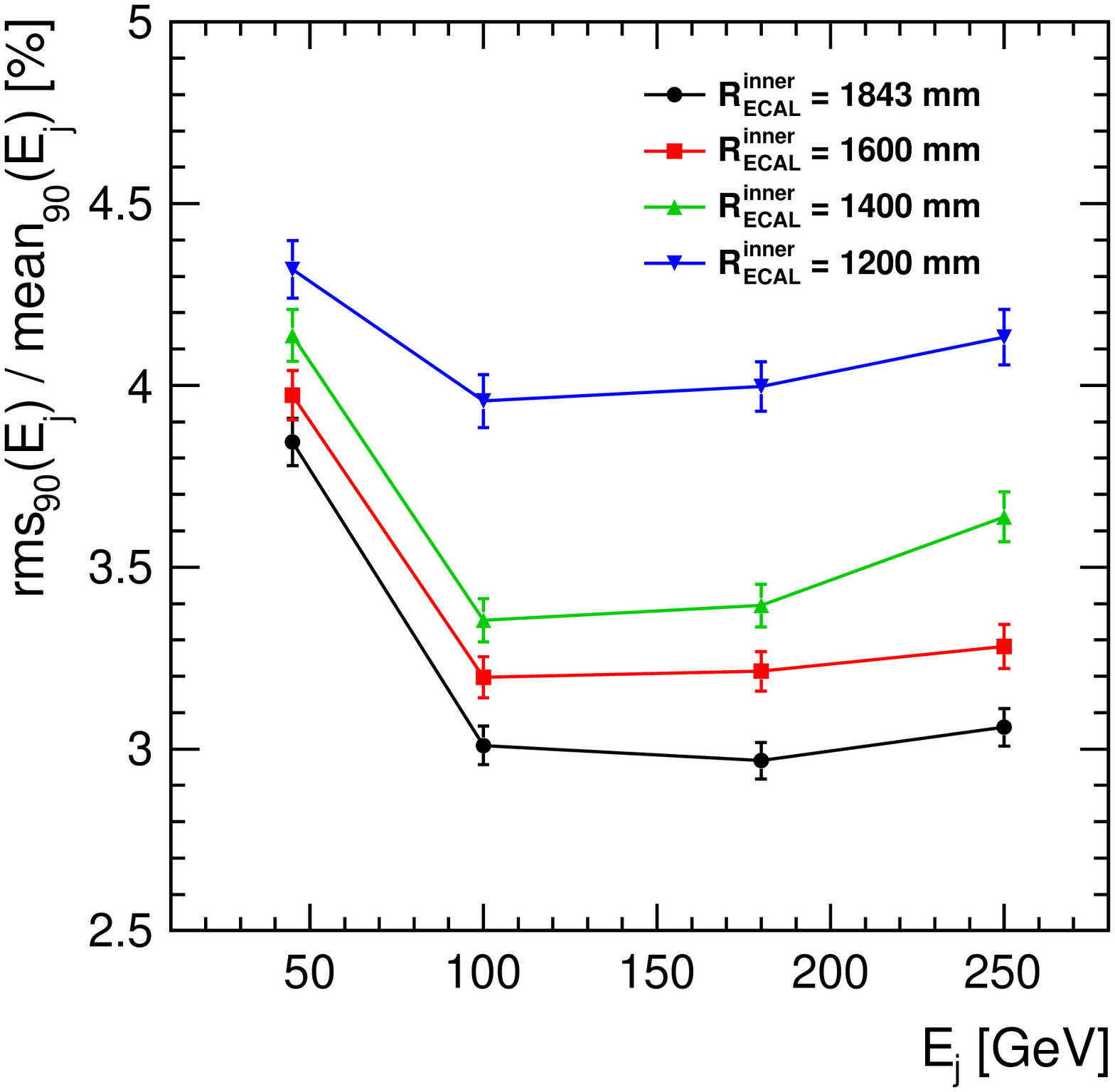}
        \caption{}
        \label{Fig:JERvsE}
    \end{subfigure}
    \caption{(a) Single jet energy resolution presented as a function of ECAL inner radius
       for jet energies between 45 and 250~GeV, and for three studies:
        SiW ECAL and AHCAL as published in the ILD LoI (dashed line), 
        a similar study with 
        recent version 0.12 of PandoraPFA (dotted lines), and SiW-ECAL+sDHCAL, with PandoraPFA version 0.09 (solid lines).     
        (b) Single jet energy resolution shown as a function of simulated jet energy for different radii.
        Only statistical errors are shown.
    }

    \label{Fig:JER}
\end{figure}
Figure~\ref{Fig:JERvsRad} shows the single jet energy resolution as a function of detector radius 
for three cases:
\begin{itemize}
    \item ILD as a combination of SiW ECAL ($5\times5$~mm$^2$ silicon cell size) and AHCAL (analog HCAL, with $3\times3$~cm$^2$ tiles) as reported in the ILD Letter of Intent~\cite{bib:ildLOI},
    \item similar study but with version 00.12 of Pandora, cf.~\cite{bib:JMecalSim},
    \item this study, where ILD consists of SiW ECAL and sDHCAL ($1\times1$~cm$^2$ cell size). The total thicknesses, cell size, number of layers of ECAL and HCAL and other parameters are kept constant.
\end{itemize}
In the first two cases, only the radius is reduced whilst in the latter case, the radius and the length of ILD are kept proportional.
The jet energy resolution degrades significantly by 11\% and 14\%, for jets at 100 and 250~GeV, correspondingly, if we choose to reduce 
the radius (and length) from 1843~mm (baseline design) to about 1400~mm.
In the first two cases and especially for jet energies at 45~GeV, there is an important difference between PandoraPFA's versions (0.09 versus 0.12), i.e. the difference in resolution between two radii $R=1.8$~m and $R=1.4$~m is 
negligible with a recent release of PandoraPFA.
For jet energies in the range of 100 to 180~GeV, the performances of 
the AHCAL and the sDHCAL are comparable at $R\sim1.4$~m. 

The jet energy resolution is also shown as a function of simulated jet energies in Figure~\ref{Fig:JERvsE} for different ILD dimensions. At different detector sizes,
the resolution behaves similarly: at low energy (45 GeV), the resolution is rather poor, this
is due to the fact that the performance is dominated by intrinsic calorimeter resolution; for jet energies above 100~GeV, the resolution starts to degrade due to the {\it confusion}, i.e. mistakes in assignment of the energies to the different reconstructed particles, especially charged and neutral particles.

From Figure~\ref{Fig:JERvsRad}, it can be seen that the jet energy resolution at $R^{\mathrm{inner}}_{\mathrm{ECAL}}=1.2$~m is significantly worse than for $R^{\mathrm{inner}}_{\mathrm{ECAL}}=1.4$~m. The latter
seems to be the minimal radius acceptable in ILD. This has to be confirmed by studies around this value, with other parameters
(number of layers in SiW ECAL, wafer size, ...)

\section{Further checks}
\label{sec:xchecks}
A few cross checks have been made to verify the results.
\subsection{Magnetic fields}
\begin{figure}
    \centering
    \begin{subfigure}[t]{0.48\textwidth}
        \includegraphics[width=\textwidth]{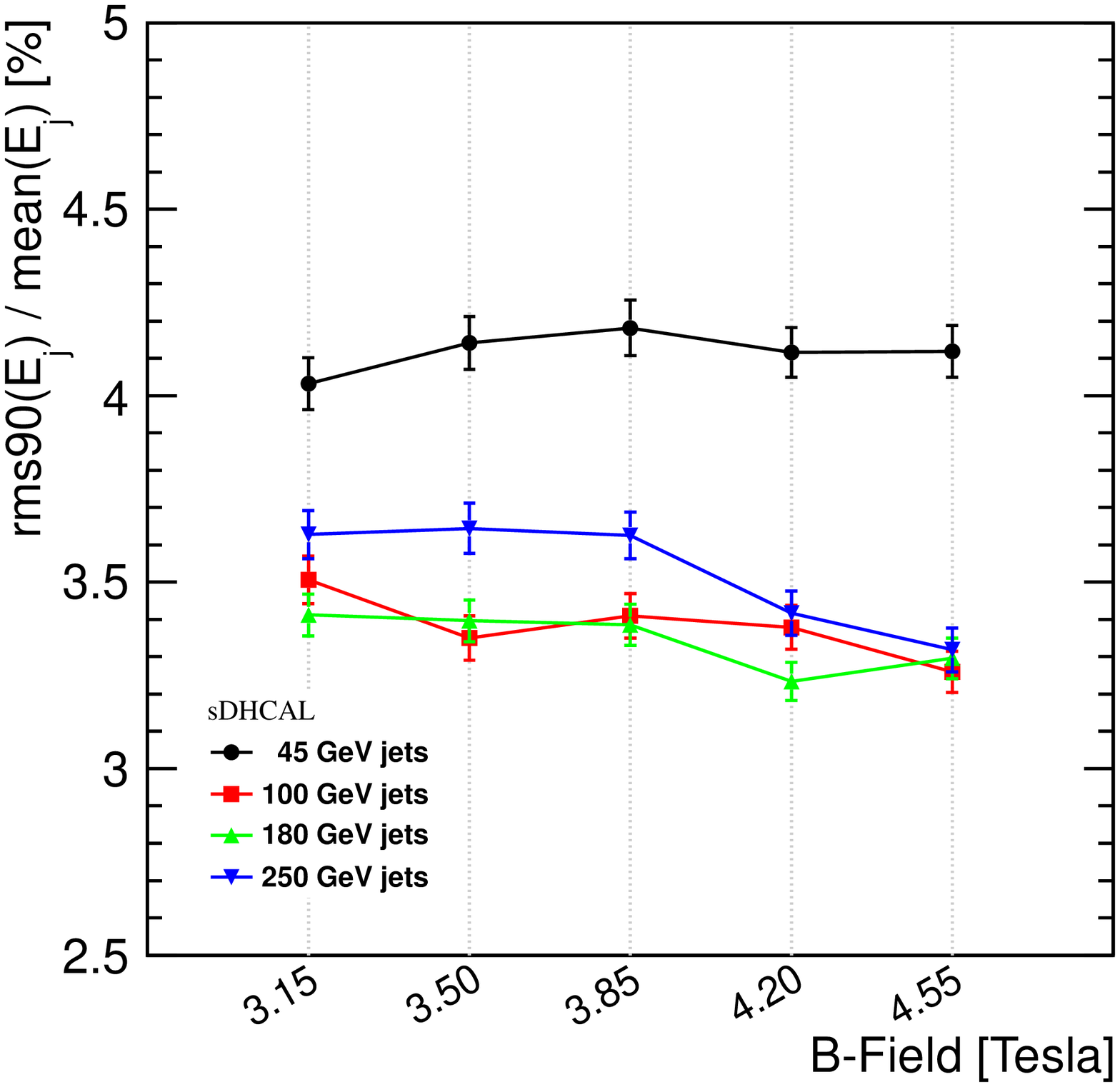}
        \caption{}
        \label{Fig:JERvsB}
    \end{subfigure}
    \quad
    \begin{subfigure}[t]{0.48\textwidth}
        \includegraphics[width=\textwidth]{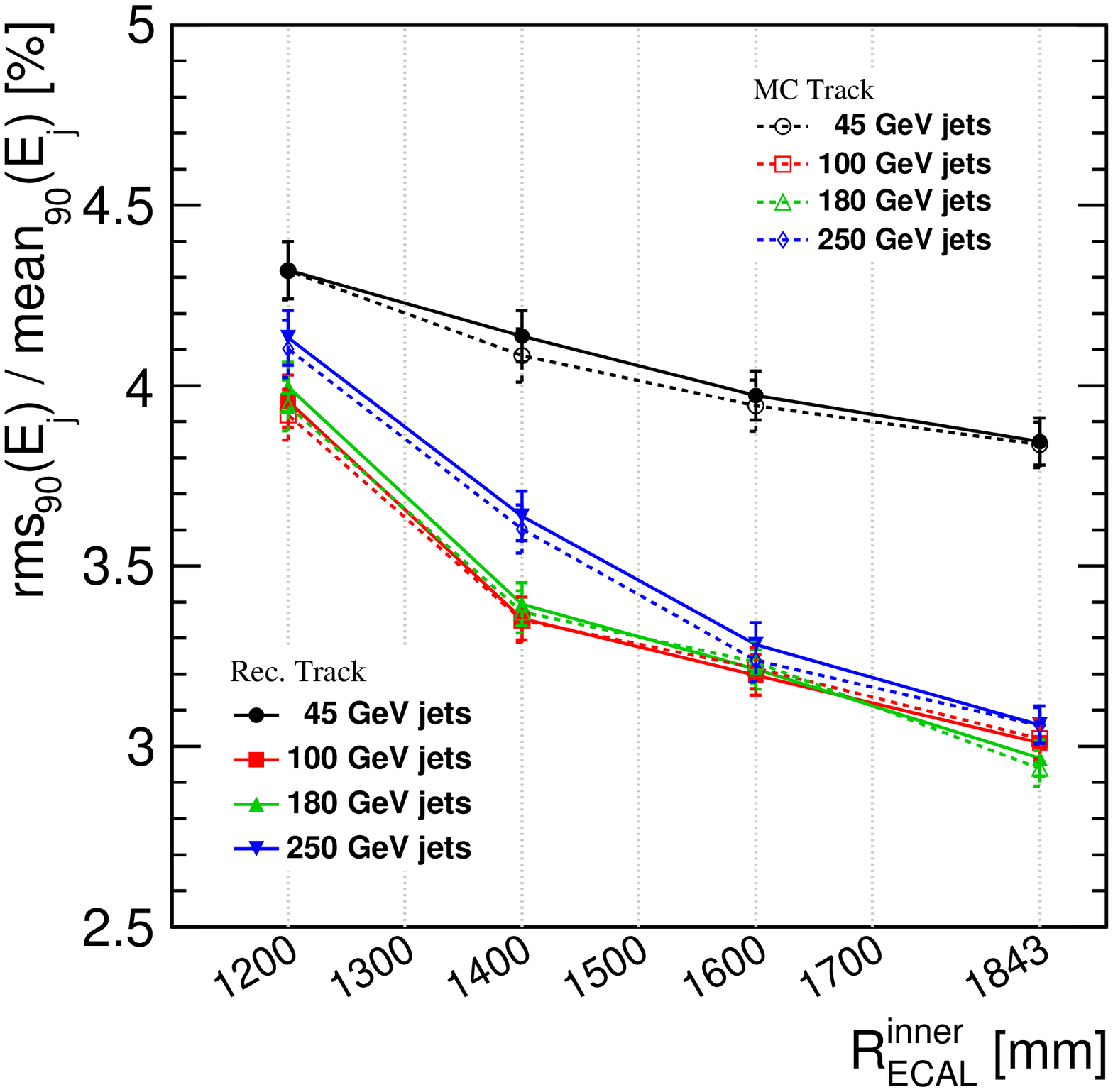}
    \caption{}
    \label{Fig:JERvsTrk}
    \end{subfigure}
    \caption{Checks of magnetic field and tracking effect on jet energy resolution.
        (a)~Single jet energy resolution as a function of magnetic field for $\ecalRinn=1.4$~m.
        (b)~Single jet energy resolution as a function of ECAL inner radius for two cases:
        with default tracking reconstruction (solid curves) and with MC truth tracks
        (dashed curves). Slight difference observed due to changes in tracking.
    }
    \label{Fig:xcheckBT}
\end{figure}

To study how the magnetic field affects jet energy resolution, we choose $\ecalRinn=1.4$~m
and vary the magnetic field $B=$ 3.15, 3.5 (default), 
3.85, 4.20 and 4.55~Tesla.
The obtained single jet energy resolution, is shown in Figure~\ref{Fig:JERvsB}.
For 45~GeV jets, for $B\geq3.5$~T, the resolution does not change significantly.
This can be explained by the fact that
the resolution at 45~GeV or less than 100~GeV is dominated by intrinsic calorimeter resolution.
As the B-field increases, there is a trade-off between the loss of low transverse
momentun tracks in the beam-pipe and improvement of the measurement of the remaining ones.
For higher energies, especially at 250~GeV, the resolution improves with B.

\subsection{Effect of tracking on jet energy resolution}
By reducing the ECAL inner radius, and hence the TPC radius, the tracking performance may degrade, it is therefore
important to quantify how it affects the PFA performance.
The comparison between 
the default reconstruction 
and the reconstruction when the track energies are taken to be the true Monte-Carlo (MC) values
are shown in Figure~\ref{Fig:JERvsTrk}. The very small difference observed demonstrates that
the track reconstruction does not affect the final Pandora performance.
\subsection{sDHCAL in analog mode}
As mentioned above, the sDHCAL sensors are GRPC and the energy estimation
is based on hit counting using semi-digital mode which helps to correct the saturation effect 
in gas. For a confirmation, the sDHCAL is simulated as an analog calorimeter, 
i.e. taking the energy deposits proportional to the charge deposits. The analysis procedure is then
performed as was done for the AHCAL. For jet at 45~GeV, we found the corresponding resolution
is about 4.23\% which is significantly worse than what is obtained by the semi-digital mode.

\section{Summary}
\label{sec:conclusions}
The relation between dimension and performance is studied for ILD using the 
combination of highest granularity calorimeters, $5\times5$~mm$^2$ cell size for the 
silicon-tungsten electromagnetic calorimeter (SiW ECAL) and $1\times1$~cm$^2$
for the semi-digital hadronic calorimeter (sDHCAL) using detailed GEANT4 models
and full reconstruction of $Z\rightarrow q\bar{q}$ events, with the $Z$ decaying at rest.

Different ILD models are investigated by varying the ECAL inner radius.
The length is modified proportionally in order to keep the same ratio as in the detailed
baseline design.  
It has been shown that the reconstructed jet energy resolution may degrade by about~8 to 19\%
for simulated jet energies in the range of~45 and~250~GeV if the radius is reduced from 1.8~m
to 1.4~m, cf. Table~\ref{tab:JERvsRad}.
Nevertheless, the models with $\ecalRinn\sim1.4$~m
still meet the ILC jet energy resolution goals, $\sigma_E/E<3.8\%$.
The price of ILD scaling roughly quadratically with the ILD dimensions may reduce by
a factor of nearly two.  It is also shown that $\ecalRinn\sim1.4$~m seems to be minimal radius acceptable for ILD.

The comparison to a similar study for the SiW ECAL and the analog hadronic calorimeter (AHCAL)
shows that the ILD performance is comparable for these two options of HCAL at $\ecalRinn\sim1.4$~m.

\begin{table}
\centerline{\begin{tabular}{ccccc}
\hline
 & \multicolumn{4}{c}{$E_{j}$ (GeV)} \\
& 45.6 & 100 & 180 & 250 \\ \cline{2-5}
 \multirow{1}{*}{$R_{\mathrm{ECAL}}^{\mathrm{inner}}$ (mm)} & \multicolumn{4}{c}{RMS$_{90}(E_j)/E_j$ (\%)} \\\hline
1843 & 3.85 & 3.01 & 2.97 & 3.06 \\\hline
1400 & 4.14 & 3.35 & 3.39 & 3.64 \\
\hline
\end{tabular}}
\caption{Jet energy resolution for different jet energies, 45.6, 100, 180, 250~GeV, and for two values of radius: 1843 and 1400~mm.}
\label{tab:JERvsRad}
\end{table}

The study on the effect of tracking on the final jet energy resolution leads to the conclusion that
the modification of ILD size may affect the tracking performance but this does not have 
significant impact on the PFA performance. It is shown that for $\ecalRinn=1400$~mm, there is possibility to increase 
slightly the magnetic field to compensate the degradation due to the reduction of the 
detector dimension, mostly at high energies (250~GeV jets).


\begin{footnotesize}
\bibliographystyle{myamsplain}
\bibliography{ildopt.bib}

\providecommand{\MR}{\relax\ifhmode\unskip\space\fi MR }
\providecommand{\MRhref}[2]{%
  \href{http://www.ams.org/mathscinet-getitem?mr=#1}{#2}
}
\providecommand{\href}[2]{#2}
\begin{thebibliography}{10}

\bibitem{bib:ildLOI}
{The ILD Concept Group}, \emph{{ILD Letter of Intent}}, DESY 2009-87, KEK
  2009-6, 2010.

\bibitem{bib:ilcTDR}
T.~Behnke {\it et al.}, \emph{{The ILC Technical Design Report}}, vol.~4,
  International Linear Collider, 2013,
  http://www.linearcollider.org/ILC/Publications/Technical-Design-Report.

\bibitem{bib:pfa1}
J.-C. Brient and H.~Videau, \emph{The calorimetry at the future $e^{+}e^{-}$
  linear collider}, Proc. of APS/DPF/DBP summer study on the future of particle
  physics (Snowmass, Colorado), 2002, http://arxiv.org/abs/hep-ex/0202004.

\bibitem{bib:pfa2}
M.~A. Thomson, \emph{{Particle Flow Calorimetry and the PandoraPFA Algorithm}},
  Nucl. Instrum. Meth. \textbf{A611} (2009), 25.

\bibitem{bib:mokka}
P.~Mora de~Freitas and H.~Videau, \emph{{Detector simulation with MOKKA /
  GEANT4: Present and future}}, Prepared for International Workshop on Linear
  Colliders (LCWS 2002), Jeju Island, Korea, 26-30, Aug 2002.

\bibitem{bib:geant4}
S.~Agostinelli {\it et al.}, \emph{{GEANT4: A Simulation toolkit}}, Nucl.
  Instrum. Meth. \textbf{A506} (2003), 250--303, SLAC-PUB-9350,
  FERMILAB-PUB-03-339.

\bibitem{bib:geant4doc}
\emph{{Geant4 Physics Reference Manual}}, Section IV, Chapter 28.

\bibitem{bib:bert}
M.P. Guthrie, R.G. Alsmiller, and H.W. Bertini, Nucl. Instrum. Meth.
  \textbf{A66} (1968), 29.

\bibitem{bib:marlin}
F.~Gaede, \emph{{Marlin and LCCD: Software tools for the ILC}}, Nucl. Instrum.
  Meth. \textbf{A559} (2006), 177--180.

\bibitem{bib:pfa3}
J.S. Marshall, A.~Muennich, and M.A. Thomson, \emph{{Performance of Particle
  Flow Calorimetry at CLIC}}, Nucl. Instrum. Meth. \textbf{A700} (2013),
  153--162.

\bibitem{bib:pandora2}
J.S. Marshall and M.A. Thomson, \emph{{Pandora Particle Flow Algorithm}}, Proc.
  of CHEF (Paris, France), April 2013.

\bibitem{bib:sdhcalcalib}
The~CALICE Collaboration, \emph{{The Time Structure of Hadronic Showers in
  Tungsten and Steel with T3B}}, CALICE Analysis Note \textbf{CAN-038} (2012).

\bibitem{bib:JMecalSim}
J.S. Marshall, \emph{{Pandora PFA with SiW and ScW ECAL models}}, International
  Workshop on Future Linear Colliders, LCWS13, Nov 2013.

\end{thebibliography}

\end{footnotesize}

\end{document}